\documentclass[aps,preprint,superscriptaddress]{revtex4-1}
\usepackage{graphicx}
\usepackage{epsfig}
\usepackage{amsmath}
\usepackage{color}
\usepackage{setspace}
\usepackage{relsize}

\renewcommand{\vec}[1]{\mathbf{#1}}
\newcommand{\fpar}[2]{\frac{{\textstyle \partial \/ #1}}{{\textstyle \partial \/ #2}}}

\begin{document}
\singlespace
\title{Hydrodynamic model for expansion and collisional relaxation of x-ray laser-excited multi-component nanoplasma}

\author{Vikrant Saxena \footnote{vikrant.saxena@desy.de}}
\affiliation{Center for Free-Electron Laser Science, Deutsches Elektronen-Synchrotron DESY, Notkestrasse 85, 22607 Hamburg, Germany.}
\affiliation{Hamburg Center for Ultrafast Imaging, Luruper Chaussee 149, 22761 Hamburg, Germany.}

\author{Beata Ziaja \footnote{ziaja@mail.desy.de}}
\affiliation{Center for Free-Electron Laser Science, Deutsches Elektronen-Synchrotron DESY, Notkestrasse 85, 22607 Hamburg, Germany.}
\affiliation{Hamburg Center for Ultrafast Imaging, Luruper Chaussee 149, 22761 Hamburg, Germany.}
\affiliation{Institute of Nuclear Physics, Polish Academy of Sciences, Radzikowskiego 152, 31-342 Krak\'ow, Poland}

\begin{abstract}
The irradiation of an atomic cluster with a femtosecond x-ray free-electron laser pulse results in a nanoplasma formation. This typically occurs within a few hundreds femtoseconds. By this time the x-ray pulse is over, and the direct photoinduced processes no longer contributing. All created electrons within the nanoplasma are  thermalized.  The nanoplasma thus formed is a mixture of atoms, electrons and ions of various charges. While expanding, it is undergoing electron impact ionization and three-body recombination. Below we present a hydrodynamic model to describe the dynamics of such multi-component nanoplasma. The model equations are derived by taking the moments of the corresponding  Boltzmann kinetic equations. We include the equations obtained, together with the source terms due to electron impact ionization and three-body recombination, in our hydrodynamic solver.  Model predictions for a test case: expanding spherical Ar nanoplasma are obtained. With this model we complete  the two-step approach to simulate x-ray created nanoplasmas,  enabling computationally efficient simulations of  their picosecond dynamics. Moreover, the hydrodynamic framework including collisional processes can be easily extended for other source terms and then applied to follow relaxation of any finite non-isothermal multi-component nanoplasma with its components relaxed into local thermodynamic equilibrium.  
\end{abstract}

\date{\today}
\maketitle

\section{Introduction} 

The recently developed x-ray free-electron lasers (FEL) \cite{FLASH,LCLS,SACLA} open up new horizons in the experimental investigation of  interaction of ultrashort intense light pulses with matter. In particular, the progressing experimental studies of  FEL irradiated  atomic clusters   \cite{Thomas2013PRL,Gorkhover2012PRL} contribute towards the understanding of  the behavior of radiation-induced ionization dynamics within complex systems. Recently, a nanoplasma formation following the irradiation of Ar cluster by a femtosecond x-ray pulse was also reported in \cite{ArClusterSACLA}. 

In case of such short irradiating pulses, one can separate the ionization dynamics of irradiated cluster into two phases. In the first (strongly non-equilibrium) evolution phase, the direct photoinduced processes including photoionization and Auger decay contribute. Electrons created within the system are not yet in thermal equilibrium. This phase lasts for up to a few hundreds fs, depending on the system size and the irradiation conditions. During this time electrons attain a local thermodynamic equilibrium. When the relaxation processes are finished, one can also assume all ions to be in their ground states. This first phase can be efficiently modeled by classical molecular dynamics (MD) while treating all scattering processes with Monte-Carlo approach \cite{xmdyn,jurek1,C60LCLS,ArClusterSACLA}. In the second evolution phase, the direct photoinduced processes do not contribute any longer, and the ionization dynamics is governed mainly by collisional processes, namely, electron impact ionization and three-body recombination. Moreover, the nanoplasma formed during the first phase continues to expand due to the thermal electron pressure and Coulomb repulsion between positively charged ions. This occurs on picosecond timescales. Therefore,  in order to understand the experimental spectroscopic data of the nanoplasma which involve final charge state distribution of ions and energy distributions of electrons and ions, one needs to follow its evolution on picosecond timescale.  This is computationally expensive with particle approaches. As reported in \cite{peltz2014}, it took $30$ days for a PIC code (run on an $80$ core shared-memory workstation 8 × Intel E7-8860)  to follow the dynamics of a spherical hydrogen cluster of  $25$ nm-radius ($\sim 2.7 \cdot 10^6$ atoms) during the first 100 fs after its irradiation. As another example, our in-house developed MD code, XMDYN \cite{xmdyn}  uses around $48$ CPU hours to simulate the dynamics of an Ar$_{1000}$ cluster for about $1.5$ ps \cite{ArClusterSACLA}. In contrast, the hydrodynamic approach (HYDRO) can achieve much better computational efficiency at simulating long-timescale evolution of such systems. Saxena et al. \cite{xhydro1}, have shown a comparison of the two approaches, MD one and hydrodynamic one, for a model system involving $ > 10^5$ particles. However, the hydrodynamic model used there enabled only sample propagation and neglected  collisional relaxation processes.

In the present work we extend the model described in \cite{xhydro1} to include impact ionization and three-body recombination processes. The extended scheme  can treat a  multi-component nanoplasma, for which electrons, neutrals (atoms) and  ground state ions are  modeled as separate fluids intermixed with each other. To achieve this, we derive the hydrodynamic equations directly from the dedicated Boltzmann equations \cite{beata_epjd, beata_prl2009} by taking their corresponding moments. The kinetic equations have been successfully applied for a description of  the initial cluster experiments performed at the FLASH facility \cite{beata_prl2009,beata_njp2009,beata_prl2010}. The three-body recombination can contribute at all stages of the sample evolution, also during the non-equilibrium phase. Its contribution depends on the transient electron density and  transient electron temperature. Following the methodology presented in \cite{flychk}, our model  accounts for  the contribution of  three-body recombination at each evolution stage, also out of equilibrium. Using the microscopic reversibility principle, the recombination is treated as an inverse process to collisional ionization. In this way the only external input required for equations are collisional cross sections. This enables, e.g., an easy  inclusion of the hydrodynamic simulation scheme into the previously discussed two-step simulation scheme (MD-HYDRO) - consistent with the parameters of the MD model used.

To compare, in a recent work by M\"uller et al.\cite{muller2009} authors have proposed source terms for fluid equations due to electron impact ionization using a similar formulation, however, their model does not include the three body recombination process.  In the two-fluid model by Meier and Schumlak \cite{meier2012}  impact ionization and radiative recombination processes are included, however, only for singly charged ions, electrons and neutrals. The dynamics of higher charge states is not treated. Three-body recombination is not treated as well. A similar scheme formulated by Khomenko et al.\cite{khomenko2014} for a multi-component partially ionized solar plasma, is also limited to singly charged ions, however, it  addresses a mixture of different atomic species in the presence of external magnetic field as it is the case for a solar plasma. This model provides a general set of transport equations, including even radiative processes, but the source terms are obtained there for a simplified two-fluid case, wherein: (i) all atoms are treated as an average neutral fluid, and (ii) all electrons together with all singly charged ionic species constitute an average charged fluid. Also, there are commercially available hydrodynamic models such as, e.g., HELIOS \cite{helios}, which include many interaction processes and various geometries. However, as  these options are built-in into the code, it is difficult to vary them (while testing, e.g., various cross section parametrization), and, in turn, to make a consistent link with the non-equilibrium MD simulations.

The objective of the present work is to formulate multi-fluid model equations which describe the dynamics of a spherical nanoplasma: (i) comprising thermalized electrons, atoms and multiply charged ions, and (ii) including collisional interactions among them.
We are aware that our model can only describe the expansion phase of the laser-cluster interaction as it does not treat direct photoinduced processes and non-thermalized electrons. This is why we foresee to adopt a two-step strategy wherein the first phase is modeled by a dedicated molecular dynamics approach \cite{xmdyn}.  With the current model we complete this two-step approach to simulate x-ray created nanoplasmas,  enabling computationally efficient simulations of  their long-timescale evolution. 

In future, this hydrodynamic framework including collisional processes can be easily extended for other source terms (e.g., describing radiative processes) and then applied to follow relaxation of any finite non-isothermal multi-component nanoplasma for which the components have relaxed into local thermodynamic equilibrium.  

In the next section, we discuss the multi-fluid equations describing the dynamics of a spherically-symmetric nanoplasma. In Sec.~III numerical results for a test system: argon nanoplasma are presented. Finally in Sec.~IV we summarize our work and discuss the future directions. 

\section{Multi-fluid model including impact ionization and three-body recombination}

Hydrodynamic equations for a system consisting of electron fluid intermixed with atomic fluid and a number of positively charged ion fluids can be derived from the corresponding kinetic equations. They read \cite{beata_epjd}:
\begin{equation}
\fpar{\rho_{_{e,i}}(\vec{r},\vec{v},t)}{t} + \vec{v}\cdot\fpar{\rho_{_{e,i}}(\vec{r},\vec{v},t)}{\vec{r}} + \frac{Z_{_{e,i}}e\vec{E}}{m_{_{e,i}}}\cdot\fpar{\rho_{_{e,i}}(\vec{r},\vec{v},t)}{\vec{v}} = S_{_{e,i}}^{^{II}}+S_{_{e,i}}^{^{TBR}}
\label{boltzmann}
\end{equation}
Here $\rho$ denotes the phase space density of particles with subscripts $\it e/i$ representing electrons/ions. Among other symbols, $\vec{r}$ and $\vec{v}$ denote position and velocity vectors, $t$ represents time, $Z$ represents the charge state (-1 for electrons, 0 for neutrals etc.), $\vec{E}$ stands for total electric field and $m$ is particle mass. On the right hand side, $S^{^{II}}$ represents the source terms due to electron impact ionization, whereas $S^{^{TBR}}$ denotes the source terms related to three body recombination. In the following, we elaborate on the source terms of Eq.(\ref{boltzmann}). The impact ionization source terms for electrons and ions are as in \cite{beata_epjd}:
\begin{eqnarray} 
S_{_e}^{^{II}}& =& \sum_{j=0}^{j_{_{max}}} n_{_j}(\vec{r},t)\biggl\{ \int d^3v_{_e} \,v_{_e} \,\rho_{_e}(\vec{r},\vec{v}_{_e},t)\left(\frac{d\sigma_{ic}^{^{j\rightarrow j+1}}(\vec{v}_{_e};\vec{v}^{\prime}_{_e}=\vec{v})}{d\vec{v}}+\frac{d\sigma_{ic}^{^{j\rightarrow j+1}}(\vec{v}_{_e};\vec{v}_{_s}=\vec{v})}{d\vec{v}} \right) \nonumber \\
&-& v\rho_{_e}(\vec{r},\vec{v},t)\sigma_{ic}^{^{j\rightarrow j+1}}(\vec{v})\biggr\}
\label{II_el}
\end{eqnarray} 
\begin{eqnarray} 
S_{_j}^{^{II}} &=& \left\{ \int d^3v_{_e} \,\sigma_{ic}^{^{j-1\rightarrow j}}(\vec{v}_{_e})\,v_{_e}\,\rho_{_e}(\vec{r},\vec{v}_{_e},t)\right\}\rho_{_{j-1}}(\vec{r},\vec{v},t) \nonumber \\
&-& \left\{ \int d^3v_{_e}\, \sigma_{ic}^{^{j\rightarrow j+1}}(\vec{v}_{_e})\,v_{_e}\,\rho_{_e}(\vec{r},\vec{v}_{_e},t)\right\}\rho_{_{j}}(\vec{r},\vec{v},t)
\label{II_ion}
\end{eqnarray} 
\noindent
Index $j$ describes a charge state of an ion and assumes values from $0$ (atoms) to $j_{_{max}}$, where $j_{_{max}}$ corresponds to the highest charge state allowed in the system. Moreover, $n_j$ is the density of ion fluid consisting of 
charge species $j$;  $v\equiv\left|\vec{v}\right|$ and 
$\sigma_{ic}^{^{j\rightarrow j+1}}(\vec{v})$ is the total impact ionization cross section for ions in the charge state $j$.  The doubly differential cross section is denoted as $d\sigma_{ic}^{^{j\rightarrow j+1}}(\vec{v}_{_e},\vec{v}^{\prime}_{_e}=\vec{v})/d\vec{v}$, where $\vec{v}_{_e}$ stands for the velocity of incoming electron and $\vec{v}^{\prime}_{_e}$ is the velocity of the incoming electron after the collision. The velocity of the secondary electron is denoted by $\vec{v}_{_s}$.
With this, the source terms describing the effect of three body recombination on the phase space density of electrons and ions can be written as in \cite{beata_epjd, beata_prl2009},
\begin{eqnarray} 
S_{_e}^{^{TBR}} = \sum_{j=0}^{j_{_{max}}} n_{_j}(\vec{r},t) \biggl\{
&& \int d^3v_{_{ep}}d^3v_{_{e}}\rho_{_e}(\vec{r},\vec{v}_{_{ep}},t)\rho_{_e}(\vec{r},\vec{v}_{_{e}},t)v_{_{ep}}v_{_e}
\frac{d\sigma_3^{^{j\rightarrow j-1}}(\vec{v}_{_{ep}},\vec{v}_{_e};\vec{v}^{\prime}_{_e}=\vec{v})}{d\vec{v}} \nonumber \\
&-& 2v\rho_{_e}(\vec{r},\vec{v},t)\int d^3v_{_{e}}d^3v^{\prime}_{_{e}}\rho_{_e}(\vec{r},\vec{v}_{_e},t)v_{_e}
\frac{d\sigma_3^{^{j\rightarrow j-1}}(\vec{v},\vec{v}_{_e};\vec{v}^{\prime}_{_e})}{d\vec{v}^{\prime}_{_e}} 
\label{TBR_el}
\end{eqnarray} 
\begin{eqnarray} 
S_{_j}^{^{TBR}} &=& 
\rho_{_{j+1}}(\vec{r},\vec{v},t) \int d^3v_{_{ep}}d^3v_{_{e}}d^3v^{\prime}_{_{e}}\rho_{_e}(\vec{r},\vec{v}_{_{ep}},t)\rho_{_e}(\vec{r},\vec{v}_{_{e}},t)v_{_{ep}}v_{_e}
\frac{d\sigma_3^{^{j+1\rightarrow j}}(\vec{v}_{_{ep}},\vec{v}_{_e};\vec{v}^{\prime}_{_e})}{d\vec{v}^{\prime}_{_e}} \nonumber \\
&-& 
\rho_{_{j}}(\vec{r},\vec{v},t) \int d^3v_{_{ep}}d^3v_{_{e}}d^3v^{\prime}_{_{e}}\rho_{_e}(\vec{r},\vec{v}_{_{ep}},t)\rho_{_e}(\vec{r},\vec{v}_{_{e}},t)v_{_{ep}}v_{_e}
\frac{d\sigma_3^{^{j\rightarrow j-1}}(\vec{v}_{_{ep}},\vec{v}_{_e};\vec{v}^{\prime}_{_e})}{d\vec{v}^{\prime}_{_e}}
\label{TBR_ion}
\end{eqnarray} 
Here $\vec{v}_{_{ep}}$ denotes the velocity of the electron which is captured during the three body recombination process, while the other electron with the velocity $\vec{v}_{_e} (\vec{v}^{\prime}_{_e})$ is the `spectator' electron. The differential cross section for the three-body recombination can be expressed in terms of doubly differential cross sections for impact ionization, using the Fowler relations for microscopic reversibility \cite{flychk}. The relation between the two differential cross sections can be written as,

\begin{equation}
\frac{d\sigma_3^{^{j\rightarrow j-1}}(v^{\prime},v'';v)}{dv} = \frac{g_{_j}}{g_{_{j+1}}} \frac{v^3}{{v^{\prime}}^3{v''}^3} \frac{h^3}{8\pi m_{_e}^3}\frac{d\sigma^{^{j-1\rightarrow j}}(v;v^{\prime})}{dv^{\prime}}mv'' \delta(E-E_{_{j-1}}-E^{\prime}-E''),
\end{equation}
where 
\begin{equation}
\delta(E-E_{_{j-1}}-E^{\prime}_{_e}-E_{_e}) = \frac{\delta\left(v^{\prime}_{_e}-\sqrt{\frac{2}{m_{_e}}(E-E_{_{j-1}}-E_{_e})}\right)}{m_{_e}\sqrt{\frac{2}{m_{_e}}(E-E_{_{j-1}}-E_{_e})}},
\end{equation}
and $E_{_{j-1}}$ is an energy needed to ionize the ion from the charge state $j-1$ to $j$. We also denote:
\begin{equation}
\widetilde{N}_{_{j-1}}\equiv \frac{h^3}{8\pi m^3_{_e}}\frac{g_{_{j-1}}}{g_{_j}} 
\end{equation}
and
\begin{equation}
f_M(\vec{v}) \equiv  (m_{_{e/i}}/2\pi k_B T_{_{e/i}})^{3/2} exp\left(-v^2/v_{_T}^2\right)
\end{equation}
which is the Maxwell-Boltzmann distribution. In the above formulae, $h$ is Planck constant, $k_B$ is Boltzmann constant, $g_{_j},  g_{_{j-1}}$ are the number of free occupancies available for an electron within an ion recombining from the charge state $j$ to the charge state $j-1$, and $v_{_T} = 2k_BT_{_{e/i}}/m_{_{e/i}}$ is the thermal velocity of electrons (ions). Substituting the above expressions in (\ref{TBR_el}) and (\ref{TBR_ion})  and factorizing the phase-space distribution of thermalized particles, $\rho(\vec{r},\vec{v},t)$, into  $\rho(\vec{r},\vec{v},t)= n(\vec{r},t)f_{M}(\vec{v}-\vec{u})$, one obtains:
\begin{eqnarray}
S_{_e}^{^{TBR}} = \sum_{j=0}^{j_{_{max}}} && n_{_j}(\vec{r},t) n_{_e}^2(\vec{r},t)4\pi \widetilde{N}_{_{j-1}}\biggl\{
vf_{M}(v)f_{M}(\hat{i}v_{_{j-1}})\sigma^{^{j-1\rightarrow j}}(v) \nonumber \\
&-& \frac{2}{v^2}f_{M}(\vec{v}-\vec{u})exp(v^2/v_{_T}^2+v_{_{j-1}}^2/v_{_T}^2)\, \int dv^{\prime}_{_e}\,{v^{\prime}_{_e}}^3\,\frac{d\sigma^{^{j-1\rightarrow j}}(v^{\prime}_{_e};v)}{dv}f_{M}(v^{\prime}_{_e})\biggr\}
\label{TBR_el2}
\end{eqnarray}
\begin{eqnarray}
S_{_j}^{^{TBR}} &=&
\rho_{_{j+1}}(\vec{r},\vec{v},t)n_{_e}^2(\vec{r},t)4\pi \widetilde{N}_{_{j}} f_{M}(\hat{i}v_{_{j}}) R^{^{j\rightarrow j+1}} \nonumber\\
&-& \rho_{_{j}}(\vec{r},\vec{v},t)n_{_e}^2(\vec{r},t)4\pi \widetilde{N}_{_{j-1}} f_{M}(\hat{i}v_{_{j-1}}) R^{^{j-1\rightarrow j}}
\label{TBR_ion2}
\end{eqnarray}
where $\hat{i}\equiv\sqrt{-1}$, and $R^{^{j\rightarrow j+1}} \equiv  \int d^3v_{_e} \,v_{_e}\, \sigma_{ic}^{^{j\rightarrow j+1}} (v_{_e})\,f_{M}(v_{_e})$ is the rate for the impact ionization from charge state $j$ to $j+1$.

Now we take zeroth, first and second order moments of Eq.(\ref{boltzmann}), along with the expressions for source terms in Eq.(\ref{II_el}) and Eq.(\ref{TBR_el2}) for electrons and those in Eq.(\ref{II_ion}) and Eq.(\ref{TBR_ion2}) for ion/atom species  respectively. Taking an $n^{th}$ order moment involves multiplying the equation by $\vec{v}^{n}$ and then integrating it over the $v$-space. We use a truncation type closure where we assume that the heat flux density vanishes, i.e., $q_{e/i} = -k_B\, \fpar{T_{e/i}}{\vec{r}}=0$ for all species \cite{book}. This enforces the uniformity of the fluid temperatures (separately for each component) within the simulation box. We make another simplifying approximation that the radial flow velocity, $u_{_e}$, of the electron fluid remains lower than its thermal velocity, i.e., $u_{_e} << v_{_T}$ which is justified in case of the slow hydrodynamic expansion of nanoplasma created after x-ray irradiation. For a spherically symmetric nanoplasma we can then write down the set of final fluid equations as:
\begin{equation}
\fpar{(r^2n_{_e})}{t}+\fpar{\left(r^2n_{_e}u_{_e}\right)}{r} = \sum_{j=0}^{j_{_{max}}} r^2 n_{_j} n_{_e} R^{^{j\rightarrow j+1}} - \sum_{j=1}^{j_{_{max}}} 4\pi r^2n_{_j}n_{_e}^2\widetilde{N}_{_{j-1}}f_{M}(\hat{i}v_{_{j-1}})R^{^{j-1\rightarrow j}}
\label{el_cont_1}
\end{equation}  
\begin{equation}
\fpar{(r^2n_{_e} u_{_e})}{t}+ \fpar{\left(r^2 n_{_e} u_{_e}^2 \right)}{r} = 
\frac{e(r^2n_{_e})}{m_{_e}}\fpar{\phi}{r} - \frac{r^2T_{_e}}{m_{_e}}\fpar{n_{_e}}{r} 
\label{el_moment_1}
\end{equation}  
\begin{eqnarray}
\fpar{T_{_e}}{t} = \frac{4\pi}{N_{_e}}\int dr \biggl[ &-& \sum_{j=0}^{j_{_{max}}}\left(T_{_e} - \frac{1}{3}m_{_e}u_{_e}^2 + \frac{2}{3} E_{_j}\right)r^2n_{_e}n_{_j}R^{^{j \rightarrow j+1}}  \nonumber \\
&+&\sum_{j=1}^{j_{_{max}}}\left(T_{_e} - \frac{1}{3}m_{_e}u_{_e}^2 + \frac{2}{3} E_{_{j-1}}\right)4\pi r^2n_{_e}^2n_{_j}\widetilde{N}_{_{j-1}}f_{M}(\hat{i}v_{_{j-1}})R^{^{j-1 \rightarrow j}} \nonumber \\
&-&  \frac{2}{3}n_{_e}T_{_e}\fpar{}{r}\left(r^2 u_{_e}\right) \biggr]
\label{Te_ode}
\end{eqnarray}  
\begin{eqnarray}
\fpar{(r^2n_{_j})}{t}+\fpar{\left(r^2n_{_j}u_{_j}\right)}{r} &=& \left(r^2n_{_e}n_{_{j-1}}-4\pi r^2n_{_e}^2n_{_j}\widetilde{N}_{_{j-1}}f_{M}(\hat{i}v_{_{j-1}})\right)R^{^{j-1 \rightarrow j}} \nonumber \\ 
&-& \left(r^2n_{_e}n_{_j}-4\pi r^2n_{_e}^2n_{_{j+1}}\widetilde{N}_{_j}f_{M}(\hat{i}v_{_{j}}) \right)R^{^{j \rightarrow j+1}}
\label{ion_cont_1}
\end{eqnarray}  
\begin{eqnarray}
\fpar{(r^2n_{_j} u_{_j})}{t}+ \fpar{\left(r^2 n_{_j} {u_{_j}}^2\right)}{r} &=& 
-\frac{Z_{_j}e(r^2n_{_j})}{m_{_i}}\fpar{\phi}{r} - \frac{r^2T_{_j}}{m_{_i}}\fpar{n_{_j}}{r}\nonumber \\
&+& \left(r^2n_{_e}n_{_{j-1}}u_{_{j-1}}-4\pi r^2n_{_e}^2n_{_i}u_{_i}\widetilde{N}_{_{j-1}}f_{M}(\hat{i}v_{_{j-1}})\right)R^{^{j-1 \rightarrow j}} \nonumber \\ 
&-& \left(r^2n_{_e}n_{_j}u_{_j}-4\pi r^2n_{_e}^2n_{_{j+1}}u_{_{j+1}}\widetilde{N}_{_j}f_{M}(\hat{i}v_{_{j}})\right)R^{^{j \rightarrow j+1}}
\label{ion_moment_1}
\end{eqnarray}  
\begin{eqnarray}
\fpar{T_{_j}}{t} = \frac{4\pi}{N_{_j}}\int dr \biggl[ & & \left\{(T_{_{j-1}}-T_{_{j}}) + \frac{1}{3}m_{_i}(u_{_{j-1}}-u_{j})^2 \right\}r^2n_{_e}n_{_{j-1}}R^{^{j-1 \rightarrow j}}  \nonumber \\
&+& \left\{(T_{_{j+1}}-T_{_j}) + \frac{1}{3}m_{_i}(u_{_{j+1}}-u_{_j})^2 \right\}4\pi r^2n_{_e}^2n_{_{j+1}}\widetilde{N}_{_j}f_{M}(\hat{i}v_{_{j}})R^{^{j \rightarrow j+1}} \nonumber \\
&-&  \frac{2}{3}n_{_j}T_{_j}\fpar{}{r}\left(r^2 u_{_j}\right) \biggr]
\label{Ti_ode}
\end{eqnarray}  
\begin{equation}
\frac{1}{r^2}\fpar{}{r} \left(r^2 \fpar{\phi}{r}\right)= \frac{e}{\epsilon_{_0}}\left(n_{_e}-\sum_{j=1}^{j_{_{max}}} Z_{_j}\,n_{_j}\right)
\label{pois}
\end{equation}

Symbols $n$, $u$, and $T$  stand for the fluid density, radial fluid velocity and fluid temperature respectively, whereas the subscripts $e$ and $j$  represent the electron fluid and ion fluid of charge state $+j$ ($j=0,\ldots,j_{max}$). This is to be noted that $u_{_{e,j}}$ represent the radial components of the flow velocities of electron(e) and ion(j) fluids and can take both positive and negative values depending on whether the flow is outward or inward.
The velocity $v_{_j}$ is the velocity of an electron of the kinetic energy corresponding to the ionization energy,  $E_{_j}$ of  the charge state  $j$ ($v_{_j} = \sqrt{2E_{_j}/m_{_e}}$).  Moreover, symbols $N_{_e}$ and $N_{_j}$ denote the total number of electrons and ions of charge $j$ respectively within the simulation box. Their values evolve with time. The electrostatic potential is denoted by the symbol $\phi$, $e$ stands for the magnitude of the electronic charge, $Z_{_j}$ is the charge of the ion fluid of the charge $j$, and $\epsilon_0$ is the vacuum permittivity.

The first equation, Eq.~(\ref{el_cont_1}), is the continuity equation for the electron fluid and describes the conservation of electron number. Eq.~(\ref{el_moment_1}) describes the conservation of the electron fluid momentum. Similarly, Eq.~(\ref{ion_cont_1}) and Eq.~(\ref{ion_moment_1}) are the continuity and momentum equations for ion fluid. Eqs.~(\ref{Te_ode}),(\ref{Ti_ode}) determine the time evolution of the electron fluid and ion fluid temperatures.
  
The spatial profile of the electrostatic potential depends on the charge density distribution. It is calculated with Poisson  equation, Eq.(\ref{pois}). The internal electric field which can be obtained with the electrostatic potential triggers the dynamics of the charged fluids, along with the thermal pressure (the second term on the right hand side of the electron and ion momentum equations) and with short-range collisions. In order to follow the nanoplasma dynamics, we solve this coupled set of  time-dependent partial differential equations, i.e.,  Eqs.  (\ref{el_cont_1}),(\ref{el_moment_1}),(\ref{ion_cont_1}),(\ref{ion_moment_1}) together with Eq. (\ref{Te_ode}), Eq. (\ref{Ti_ode}) and Eq. (\ref{pois}). In order to demonstrate the complex dynamics, described by these equations, in the next section we will apply them to a simple study case - preheated argon nanoplasma.


\section{Numerical simulation of  expanding argon nanoplasma}


The set of equations presented in Sec.~II is an extended version of the model presented in \cite{xhydro1}. The inclusion of  the new source terms representing the impact ionization and three body recombination of these collisional processes in our previous model does not change the basic properties of these equations. We can therefore use the same numerical methods as in \cite{xhydro1} to solve the new set of equations: (i) the flux corrected transport (FCT) scheme \cite{lcpfct,boris} for the 0th and 1st moment equations, (ii) $4^{th}$-order Runge-Kutta integration \cite{num_recipe} for the fluid temperature equations, and (iii) tri-diagonal method \cite{num_recipe} for the Poisson's equation. Moreover, to ensure the stability against sharp gradients, we use the artificial viscosity concept introduced by Lapidus \cite{lapidus,lohner}. 
 
As a test case we choose a quasi-neutral spherical nanoplasma of radius $100$ \AA$\,$ which consists of a positively charged fluid of $Ar^{+1}$ ions at room temperature ($T_{+1}=0.025$ eV at $0$ fs) intermixed with a warm electron fluid ($T_{e}=25$ eV at $0$ fs). The density of the argon nanoplasma is equal to $1.9742\times 10^{28}$ m$^{-3}$ which corresponds to an inter-atomic separation of $3.7$ \AA. For all simulation runs presented here we have considered a simulation box size of $R_{max} = 1000$ \AA$\,$ with the grid resolution fixed to  $dr = 0.1$ \AA. The initial time step is taken as $dt = 0.01$ attoseconds. Later the adaptive time step scheme is applied, following the Courant stability criterion \cite{num_recipe}. As the left boundary of our simulation box corresponds to the symmetry axis at $r=0$, we use there the reflective boundary condition. An outflow boundary condition is used at the right (external) boundary ($r=R_{max}$). The moment equations, Eqs.~(\ref{el_cont_1}), (\ref{el_moment_1}), (\ref{ion_cont_1}), (\ref{ion_moment_1}) are then numerically solved,  along with the temperature equations, Eqs.~(\ref{Te_ode}),(\ref{Ti_ode}), and the Poisson equation, Eq.~(\ref{pois}). In this study we use the parametrization of the total impact ionization cross section as given by Lotz \cite{lotz}. The rates for electron impact ionization at electron temperatures between $0 - 50$ eV are tabulated in an input file. Their intermediate values are calculated using linear interpolation. In dedicated test runs we have checked that the particle number, charge and total energy are accurately conserved during the simulations.

\begin{figure}[!h]
\includegraphics[scale=0.6,angle=0]{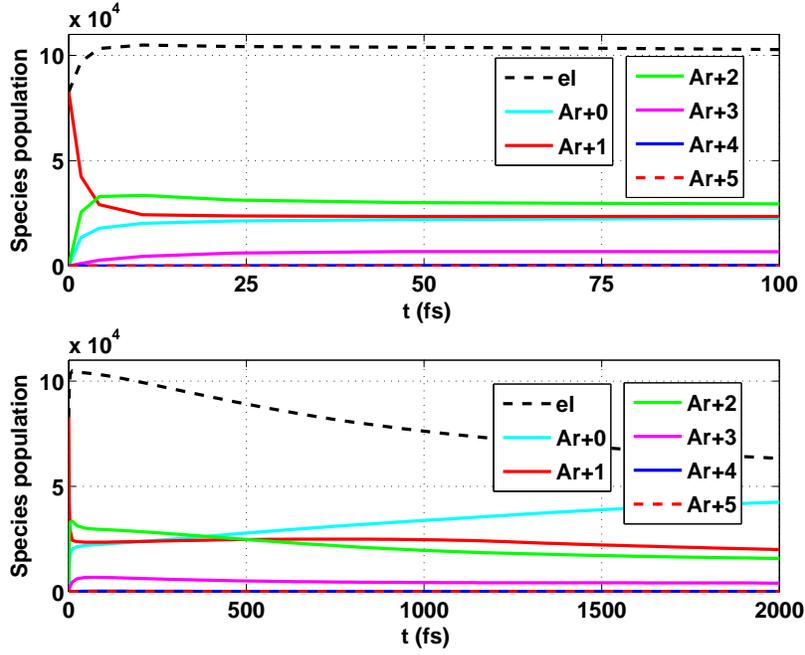}
\caption{(color online) Evolution of  electron and ion populations at times:
(a) up to 100 fs and (b) up to 2 ps. Initial fluid temperatures were: $T_{e}=25$ eV, $T_{+1}=0.025$ eV  at $0$ fs.}
\end{figure}

\begin{figure}[!h]
\includegraphics[scale=0.6,angle=0]{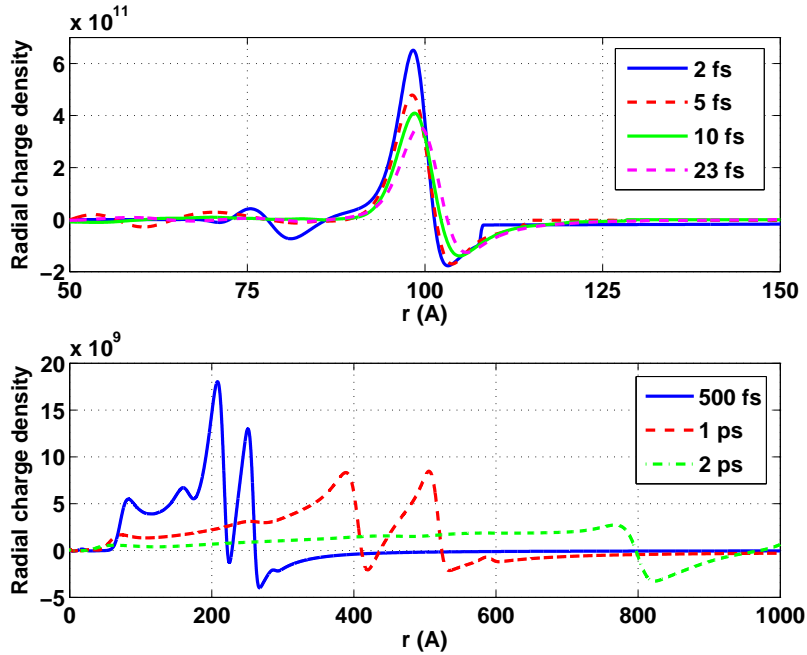}
\caption{(color online) The net charge density distribution: (a) in the initial phase of expansion, $t< 23$ fs, and   (b) at later times : $t_1=500$ fs, $t_2=1$ ps and $t_3=2$ ps.}
\end{figure}

The simulation results are discussed below. 
In Fig.~1 we show how the populations of different species evolve. Their initial temperatures were:  $T_{e}= 25$  eV and $T_{+1}= 0.025$  eV. The upper subplot shows the population evolution during the first 100 fs, and the lower one shows the population evolution for up to $2$ ps. In the initial phase of the evolution, the impact ionization dominates over three body recombination. This is because of the high temperature of electrons. At later stages, as the electron temperature decreases, three body recombination becomes more significant and the population of neutral species increases. We notice that throughout the simulation time, the populations of $+4$ and $+5$ charge states remain considerably low. During the first few hundreds femtoseconds doubly charged ions have the largest population among the positively charged species, whereas at later stages singly charged state becomes the predominant charge state, apart from the neutral species which have the highest population. 

In the first subplot of Fig.~2, the spatial distribution of  net charge density is  shown at four different time instants $2, 5, 10$ and $23$  fs during the initial phase of the nanoplasma evolution. In the lower subplot, we show the net charge density distribution after $0.5, 1$ and $2$ ps. It should be noted that the nanoplasma remains quasi-neutral in the cluster core region during the initial phase, whereas a space-charge separation develops at the cluster edge. It then keeps evolving with time, with fractions of fastest ions continuosly leaving the box.  Electron escape is even faster: a part of electron population is already moving towards the box boundary at $\sim 2$ fs. At later times the space charge separation is no longer localized and it spreads over a large region.

\begin{figure}[!h]
\includegraphics[scale=0.6,angle=0]{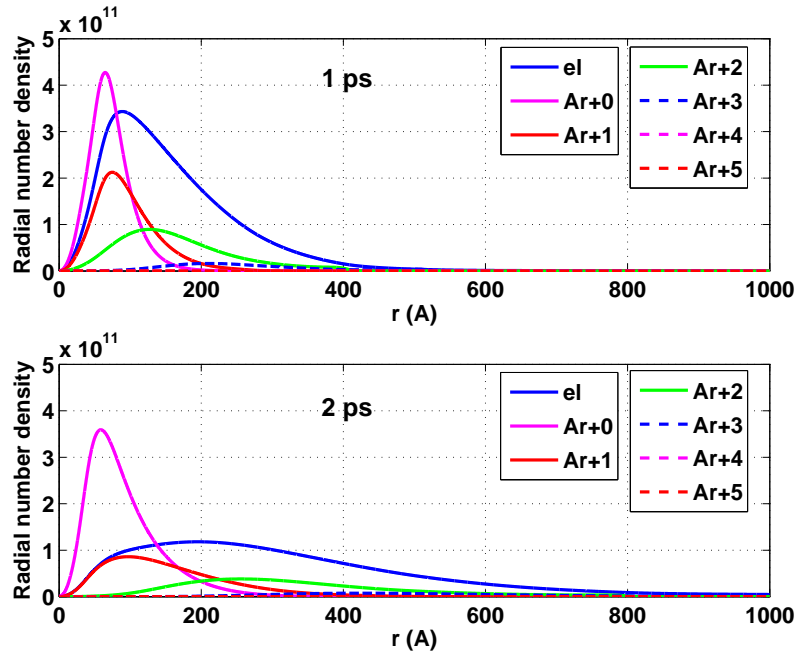}
\caption{(color online) The number density distribution of different species at: (a) 1 ps and (b) 2 ps.}
\end{figure}

\begin{figure}[!h]
\includegraphics[scale=0.6,angle=0]{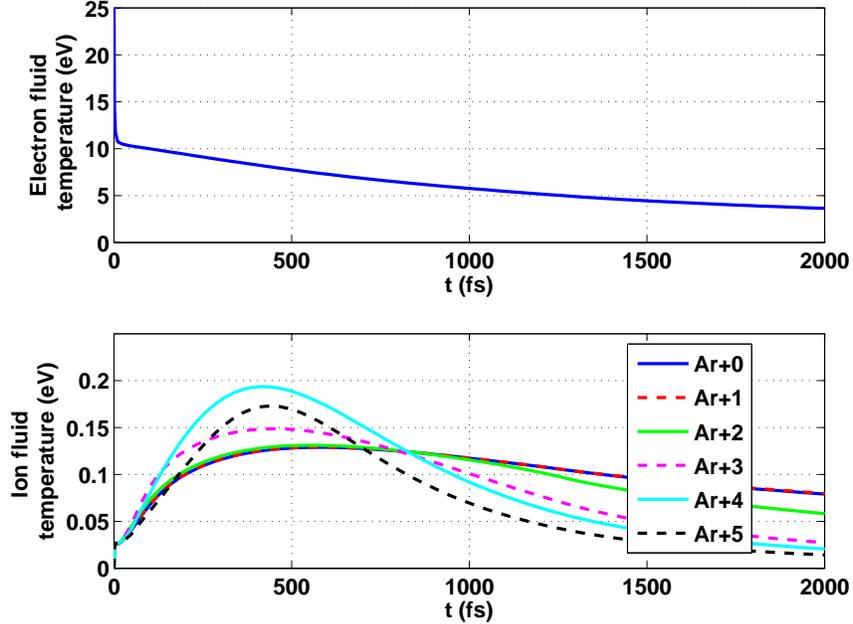}
\caption{(color online) Time evolution of the temperatures of:  (a) electron fluid (b) ion fluids.}
\end{figure}

The number densities of different species are shown in Fig.~3, at 1 ps (upper subplot) and at 2 ps  (lower subplot). The growing population of atoms in the core region at the later times indicates the predominant contribution of three body recombination due to trapped slow electrons. Moreover, in our simulations we observe that the higher charge species tend to move outwards with higher radial velocities. They eventually populate the outer region near the surface of the nanoplasma before moving out of the simulation box, as also predicted by, e.g., Siedschlag and Rost \cite{siedschlag}. This fast escape of highly charged ions contributes additionally to the increasing relative participation of neutrals in the overall charge distribution. 

Finally, in Fig.~4 we show time evolution of fluid temperatures of different species. The electron temperature, shown in the upper subplot, first decreases rapidly. This indicates that most energetic electrons escape from the box quickly. 
Further, two more processes begin to contribute. These are the electron impact ionization and thermal expansion of the nanoplasma.  They lead to the temperature decrease. We have verified that the electron fluid velocity remain much lower than the electron thermal velocity throughout the expansion of the nanoplasma, thus fulfilling the condition  ($u_e<< v_{th}$) used for deriving our model in Sec.~II.

In the lower subplot of Fig.~4, time evolution of  ion temperatures is shown.  The temperature (i.e., kinetic energy) first increases from its initial value because of the mutual repulsive Coulomb interaction between ions and their acceleration.  Consequently, this increase is stronger for highly charged ions. The temperature decreases as the fluid expands. We can see that the temperature of atoms and Ar$^{+1}$ ions remain almost identical (overlapping curves), as the atoms appear only as a result of the recombination of Ar$^{+1}$ ions.  In general, ion temperatures remain far below the electron temperature, due to the significant mass difference between electrons and ions.

The simulated evolution of the Ar nanoplasma is in agreement with our expectations. Timescales of collisional processes, their interplay in time and, finally, the expansion dynamics agree qualitatively with observations from previous simulations and experiments \cite{Thomas2013PRL,ArClusterSACLA,beata_prl2009,xhydro1}. To emphasize, the hydrodynamics simulations were performed on a single CPU for a large Ar cluster ($>80000$ atoms) and took around $24$ hrs for $1$ ps simulation. This  is significantly faster than the previously quoted MD calculations which - to  remind - took $48$ hrs on a single CPU for a $1.5$ ps long simulation of Ar$_{1000}$ cluster. These findings indicate the consistency and efficiency of the approach proposed and encourage its future application for quantitative interpretation of experimental data. 
 

\section{Summary and Outlook}


We proposed a concise multi-fluid model, dedicated to follow an expansion of a nanoplasma, formed during the interaction of an x-ray FEL pulse with an atomic cluster. The model includes collisional processes: electron impact ionization and three body recombination. The resulting stable and computationally efficient numerical scheme has been applied to a test case of argon nanoplasma. The predictions obtained were in agreement with the qualitative expectations, based on observations from previous simulations and experiments. In particular, nanoplasma remained quasi-neutral within the cluster core during the initial evolution phase, whereas a space-charge separation developed at  the cluster edge. The nanoplasma then kept evolving with time, with fractions of fastest ions continuosly leaving the box. Electron escape was very fast: large fraction of electrons  moved towards the box boundary already at $\sim 2$ fs. At later times the space-charge separation was no longer localized and spread over a large region. Quasi-neutral core region then significantly shrank. The on-going recombination and the escape of highly charged ions contributed to the increasing relative participation of neutrals in the overall  charge distribution. The dynamics of charged fluids was reflected by the corresponding changes of their temperatures. With this collisional model we completed the two-step approach to simulate x-ray created nanoplasmas,  enabling computationally efficient simulations of  their long-timescale (picosecond) dynamics. The universal hydrodynamic framework proposed can also be easily extended for other specific source terms (e.g., those describing radiative processes) and then applied to follow the evolution of any finite non-isothermal multi-component nanoplasma with its components relaxed into local thermodynamic equilibrium.  


\section{Acknowledgments}

We would like to thank Hyun-Kyung Chung, Zoltan Jurek and Robin Santra for  helpful discussions.


\begin{thebibliography}{99}



\bibitem{FLASH}
W. Ackermann, G. Asova, V. Ayvazyan, A. Azima, N. Baboi, J. B\"ahr, V. Balandin, B. Beutner, A. Brandt, A. Bolzmann {\it et al.}, Nat. Photon. {\bf 1}, 336 (2007).

\bibitem{LCLS}
P. Emma, R. Akre, J. Arthur, R. Bionta, C. Bostedt, J. Bozek, A. Brachmann, P. Bucksbaum, R. Coffee, F.-J. Decker {\it et al.}, Nat. Photon. {\bf 4}, 641 (2010).

\bibitem{SACLA}
T. Ishikawa, H. Aoyagi, T. Asaka, Y. Asano, N. Azumi, T. Bizen, H. Ego, K. Fukami, T. Fukui, Y. Furukawa {\it et al.}, Nat. Photon. {\bf 6}, 540 (2012).

\bibitem{Thomas2013PRL}
H. Thomas, A. Helal, K. Hoffmann, N. Kandadai, J. Keto, J. Andreasson, B. Iwan, M. Seibert, N. Timneanu, J. Hajdu {\it et al.}, Phys. Rev. Lett. {\bf 108}, 133401 (2012).

\bibitem{Gorkhover2012PRL}
T. Gorkhover, M. Adolph, D. Rupp, S. Schorb, S.W. Epp, B. Erk, L. Foucar, R. Hartmann, N. Kimmel, K.-U. K\"uhnel {\it et al.}, Phys. Rev. Lett. {\bf 108}, 245005 (2012). 

\bibitem{ArClusterSACLA}
T. Tachibana,  Z. Jurek, H. Fukuzawa, K. Motomura, K. Nagaya, S. Wada, P. Johnsson, M. Siano, S. Mondal, Y. Ito {\it et al.}, Sci. Rep. {\bf 5}, 10977 (2015).

\bibitem{xmdyn}
Z. Jurek, B. Ziaja, and R. Santra, XMDYN Rev. 1.0360 (CFEL, DESY, 2013).

\bibitem{jurek1}
Z. Jurek, G. Oszlanyi, and G. Faigel, Europhys. Lett. {\bf 65}, 491 (2004).

\bibitem{C60LCLS}
B.F. Murphy, T. Osipov, Z. Jurek, L. Fang, S.-K. Son, M. Mucke, J.H.D. Eland, V. Zhaunerchyk, R. Feifel, L. Avaldi {\it et al.}, Nat. Comm. {\bf 5}, 4281 (2014). 

\bibitem{peltz2014}
C. Peltz, C. Varin, T. Brabec, and T. Fennel, Phys. Rev. Lett. {\bf 113}, 133401 (2014).

\bibitem{xhydro1}
V. Saxena, Z. Jurek, B. Ziaja, and R. Santra, High Energ. Dens. Phys. {\bf 15}, 93-98 (2015).

\bibitem{beata_epjd}
B. Ziaja, A.R.B. de Castro, E. Weckert, and T. M\"oller, Eur. Phys. J. D {\bf 40}, 465-480 (2006).

\bibitem{lotz}
W. Lotz, Zeitschrift f\"ur Physik {\bf 206}, 205-211 (1967).

\bibitem{muller2009}
S.H. M\"uller, C. Holland, G.R. Tynan, J.H. Yu, and V. Naulin, Plasma Phys. Control. Fusion {\bf 51}, 105014 (2009).

\bibitem{meier2012}
E.T. Meier and U. Shumlak, Phys. Plasm. {\bf 19}, 072508 (2012).

\bibitem{khomenko2014}
E. Khomenko, M. Collados, A. Diaz, and N. Vitas, Phys. Plasm. {\bf 21}, 092901 (2014), 092901 (2014).

\bibitem{helios}
J.J. MacFarlane, I.E. Golovkin, and P.R. Woodruff, J. Quant. Spectrosc. Radiat. Transf. {\bf 99}, 381 (2006).

\bibitem{beata_prl2009}
B. Ziaja, H. Wabnitz, F. Wang, E. Weckert, and T. M\"oller, Phys. Rev. Lett. {\bf 102}, 205002 (2009).

\bibitem{beata_njp2009}
B. Ziaja, T. Laarmann, H. Wabnitz, F. Wang, E. Weckert, C. Bostedt, and T. M\"oller, New J. Phys. {\bf 11}, 103012 (2009). 

\bibitem{beata_prl2010}
R.R. F\"austlin, Th. Bornath, T. D\"oppner, S. D\"usterer, E. F\"orster, C. Fortmann, S.H. Glenzer, S. G\"ode, G. Gregori, R. Irsig {\it et al.}, Phys. Rev. Lett. {\bf 104}, 125002 (2010).

\bibitem{flychk}
H.-K. Chung, M.H. Chen, W.L. Morgan, Y. Ralchenko, and R.W. Lee, High Energ. Dens. Phys. {\bf 1}, 3-12 (2015).

\bibitem{book}
R. Fitzpatrick, {\it Plasma Physics : An Introduction} (CRC Press, Taylor and Francis Group, 2014).

\bibitem{lcpfct}
J.P. Boris, A.M. Landsberg, E.S. Oran, and J.H. Gardner, {\em LCPFCT-A flux-corrected transport algorithm for solving 
generalized continuity equations, Naval Research Lab Washington DC}, NRL/MR/6410--93-7192 (1993). 
\bibitem{boris}
J.P. Boris and D.L. Book, Meth. Comp.  Phys. {\bf 16}, 85 (1976).

\bibitem{num_recipe}
W.H. Press, S.A. Teukolsky, W.T. Vetterling, and B.P. Flannery {\em Numerical Recipes in Fortran (Cambridge University Press, 
New York, 1992)}.

\bibitem{lapidus}
A. Lapidus, J. Comp. Phys. {\bf 2}, 154-177 (1967).

\bibitem{lohner}
R. Lohner, K. Morgan, and J. Peraire, Commun.  Appl.  Num.  Meth. {\bf 1}, 141-147 (1985).

\bibitem{siedschlag}
C. Siedschlag and J.-M. Rost, Phys. Rev. Lett. {\bf 93}, 043402 (2004).
\end{thebibliography}
\end{document}